\documentclass[epj]{svjour}
\usepackage{amsmath,amssymb}
\usepackage{bm}
\usepackage{epsfig}

\newcommand{\bd}{\bm}

\begin{document}

\title{What are spin currents in Heisenberg magnets?}

\author{Florian Sch\"{u}tz\inst{1}\and Peter Kopietz\inst{1}\and Marcus Kollar\inst{2}}

\institute{Institut f\"{u}r Theoretische Physik, Universit\"{a}t
  Frankfurt, Robert-Mayer-Strasse 8, 60054 Frankfurt, Germany
  \and
  Theoretische Physik III, Elektronische Korrelationen und
  Magnetismus, Institut f\"{u}r Physik, Universit\"{a}t Augsburg,
  86135 Augsburg, Germany}

\date{May 14, 2004}

\abstract{ 
  We discuss the proper definition of the spin current operator in
  Heisenberg magnets subject to inhomogeneous magnetic fields.  We
  argue that only the component of the naive ``current operator''
  $J_{ij}{\bd S}_i\times{\bd S}_j$ in the plane spanned by the local
  order parameters $\langle{\bd S}_i\rangle$ and $\langle{\bd
    S}_j\rangle$ is related to real transport of magnetization.
  Within a mean field approximation or in the classical ground state
  the spin current therefore vanishes.  Thus, finite spin currents are
  a direct manifestation of quantum correlations in the system.
}


\maketitle

\section{Introduction}

In a recent Letter \cite{Schuetz03} and a subsequent paper
\cite{Schuetz04} we have calculated the persistent spin currents in
mesoscopic Heisenberg rings subject to inhomogeneous magnetic fields.
We have emphasized the close analogy between this phenomenon and the
well known persistent charge currents in mesoscopic metal rings
pierced by an Aharonov-Bohm flux.  In the ensuing discussions with
several colleagues we have become aware of the fact that the
definition of the spin current operator in Heisenberg magnets subject
to inhomogeneous magnetic fields is not obvious.  In this note we
shall attempt to clarify this point.  

A related problem, which will not be discussed in this work, is the
definition of the spin current operator in semi-conducting electronic
systems with strong spin-orbit interactions. Recently, Rashba
\cite{Rashba03} pointed out that also for this case the precise
meaning of the concept of a spin current is rather subtle.  In
particular, he emphasizes that spin currents in thermodynamic
equilibrium, which arise with the standard definition of the
spin-current operator used in the literature, are unphysical and
should be regarded as background currents which do not correspond to
real transport of magnetization.  A clear understanding of this
concept is essential for the highly active field of information
processing using spin degrees of freedom subsumed under the name of
spintronics \cite{Awschalom02}.
  
For itinerant systems the spin is an intrinsic property of the charge
carriers and is carried around with their motion. For localized spin
systems considered here, transport of spin is a consequence of the
time evolution of the magnetization. For special cases the transport
can be ascribed to the movement of quasi-particles as magnons or
spinons and again a simple physical picture emerges \cite{Meier02}.
  
In this context, it is also interesting to note that in effective
low-energy models for ferromagnets involving only the spin degrees of
freedom even the concept of the linear momentum is not well defined
\cite{Volovik87}. In general, the dynamical equation for the spin
degrees of freedom have to be supplemented by kinetic equations for
the underlying fermionic excitations.

At the heart of the ambiguities involved in defining a spin current
operator both for itinerant systems with spin orbit interaction as
well as for Heisenberg magnets in inhomogeneous fields is the fact
that the magnetization is not strictly conserved for these systems.
Still, the intuitive concept of magnetization transport should also be
useful for these systems and one is therefore led to define effective
current operators, as we will do in this note for the case of a
Heisenberg magnet in an inhomogeneous magnetic field.  

\section{Problems with the naive definition of the spin current operator}

\begin{figure}[tb]
  \begin{center}
    \epsfig{file=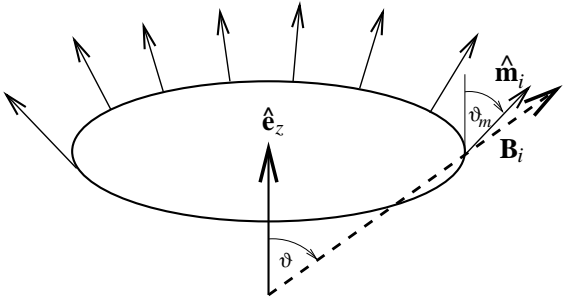,width=75mm}
  \end{center}
  \vspace{-4mm}
  \caption{%
    Classical spin configuration $\hat{\bd{m}}_i$ of a
    nearest-neighbor ferromagnetic Heisenberg ring in a radial
    magnetic field ${\bd{B}}_i$.}
  \label{fig:angle}
\end{figure}

Consider a general Heisenberg magnet with Hamiltonian
\begin{equation}
  \hat{H} = \frac{1}{2} \sum_{  i,j}   J_{ij} {\bd{S}}_i \cdot
  {\bd{S}}_j   - g \mu_{\text{B}} \sum_{i  }  {\bd{B}}_i \cdot
  {\bd{S}}_i
  \,,
  \label{eq:Hamiltonian}
\end{equation} 
where the sums are over all sites ${\bd{r}}_i$ of a chain with
periodic boundary conditions, the $J_{ij}$ are general exchange
couplings, and ${\bd{S}}_i$ are spin-$S$ operators normalized such
that ${\bd{S}}_i^2 = S ( S+1 )$.  The last term in
Eq.~(\ref{eq:Hamiltonian}) is the Zeeman energy associated with an
inhomogeneous magnetic field ${\bd{B}}_i = {\bd{B}} ( {\bd{r}}_i )$.
We assume that the magnetic field at each lattice site is sufficiently
strong to induce permanent magnetic dipole moments ${\bd{m}}_i = g
\mu_{\text{B}} \langle {\bd{S}}_i \rangle$, not necessarily parallel
to $\bd{B}_i$, where $\langle \ldots \rangle$ denotes the usual
thermal average.  The simplest geometry is a ferromagnetic ring in a
crown-shaped magnetic field, as illustrated in Fig.~\ref{fig:angle}.
This geometry is used in the following for illustrative purposes, but
our arguments are not restricted to this case.
The Hamiltonian (\ref{eq:Hamiltonian}) implies the equation of motion
\begin{equation}
  \hbar \frac{ \partial   {\bd{S}}_i}{\partial t }    
  +   {\bd{h}}_i
  \times  {\bd{S}}_i +
  \sum_{j }  {J_{ij}} {\bd{S}}_i \times {\bd{S}}_j
  = 0
  \,,
  \label{eq:eom}
\end{equation}
where ${\bd{h}}_i = g \mu_{\text{B}} {\bd{B}}_i$.  Note that the last
term in Eq. (\ref{eq:eom}) can be written in the form $\sum_j
{\bd{I}}_{i \rightarrow j}$, with
\begin{equation}
  {\bd{I}}_{i \rightarrow j} = J_{ ij}  {\bd{S}}_i \times {\bd{S}}_j
  \; .
  \label{eq:spincur1}
\end{equation}
By analogy with the discrete lattice version of the equation of
continuity for charge currents, it is tempting to identify $
{\bd{I}}_{i \rightarrow j}$ with the operator whose expectation value
gives the spin current from site $i$ to site $j$.  In this work we
shall argue that this identification is only correct for a strong
{\it{homogeneous}} magnetic field, where in equilibrium the
expectation values $ \langle {\bd{S}}_i \rangle$ of the spins at all
sites are aligned along the same spatially constant direction of the
field.  Then ${\bd{h}}_i \times \langle {\bd{S}}_i \rangle = 0$.
Using the fact that equilibrium averages are time independent,
$\frac{d}{dt} \langle {\bd{S}}_i \rangle =0$, we conclude from the
equation of motion (\ref{eq:eom}) that the lattice divergence of the
spin current in the presence of a homogeneous magnetic field vanishes
\begin{equation}
  \sum_j \langle  {\bd{I}}_{i \rightarrow j} \rangle = 0
  \; .
\end{equation}
For a one-dimensional ring with nearest neighbor hopping this implies
for each site $i$
\begin{equation}
  \langle {\bd{I}}_{ i \rightarrow i+1}\rangle  
  + \langle {\bd{I}}_{ i \rightarrow i -1} \rangle 
  =
  \langle {\bd{I}}_{ i \rightarrow i+1}\rangle  
  - \langle {\bd{I}}_{ i-1 \rightarrow i } \rangle =0 
  \; ,
\end{equation}
so that the same spin current $ \langle {\bd{I}} \rangle = \langle
{\bd{I}}_{ i \rightarrow i+1} \rangle $ flows through each link of the
ring.  However, the equation of motion contains only the divergence of
the current, so that it does not fix the value of $\langle {\bd{I}}
\rangle $.  From the point of view of elementary vector analysis this
is a consequence of the fact that both the divergence and the curl are
necessary to uniquely specify a vector field. Because the equation of
motion contains only the divergence, circulating spin currents cannot
be calculated using the equation of motion.  In fact, even the
definition of the spin current operator in a geometry permitting
circulating spin currents cannot be deduced from the equation of
motion.  Of course, for a ring with a collinear spin configuration we
know that $ \langle {\bd{I}} \rangle = 0$, so that there are no
circulating currents.

The case of a non-uniform magnetic field is more interesting.  In
general, the spin configuration in the ground state is then also
inhomogeneous. For example, let us consider a radial magnetic field
${\bd{B}}_i = |\bd{B}| {\bd{r}}_i / | {\bd{r}}_i |$ situated at
constant latitude $\vartheta_i = \vartheta$, as shown in
Fig.~\ref{fig:angle}.  We assume that the direction $\hat{\bd{m}}_i$
$=$ ${\bd{m}}_i/ | {\bd{m}}_i |$ of the magnetic moments $ {\bd{m}}_i
= g \mu_B \langle {\bd{S}}_i \rangle$ trace out a finite solid angle
$\Omega$ on the unit sphere in order-parameter space as we move once
around the ring.  If we consider a nearest neighbor Heisenberg
ferromagnet with $J_{ij} = - J ( \delta_{ i, j+1} + \delta_{ i ,
  j-1})$ then for $ | {\bd{h}} | \equiv g \mu_{\text{B}} |\bd{B}|
\gtrsim J S (2 \pi / N )^2$ the classical ground-state configuration
$\hat{\bd{m}}_i$ is radial as well, with a slightly different latitude
$\vartheta_m$ satisfying \cite{Schuetz03}
\begin{equation}
  \sin ( \vartheta_m - \vartheta ) = - ({JS}/| {\bd{h}} | ) \left[ 
    1 - \cos ( 2 \pi /N  ) \right] \sin ( 2 \vartheta_m )
  \,.
  \label{eq:classicalangle}
\end{equation} 

The main point of this work is that in the presence of an
inhomogeneous magnetic field the spin current operator is {\it{not}}
simply given by Eq. (\ref{eq:spincur1}).  The fact that the
expectation value of Eq. (\ref{eq:spincur1}) cannot be a spin current
is perhaps most obvious if we consider the simple case of classical
spins in a star-shaped magnetic field, corresponding to $\vartheta_m =
\vartheta = \pi /2$ in Fig. \ref{fig:angle}.  In this case Eq.
(\ref{eq:spincur1}) gives for a ring with evenly spaced sites at zero
temperature
\begin{equation}
  {\bd{I}}_{i \rightarrow j} = J_{ij} {\bd{e}}_z \sin ( 2 \pi / N )
  \; , 
\end{equation}
where ${\bd{e}}_z$ is a unit vector perpendicular to the plane of the
ring.  Note that at the classical level the statics and dynamics of a
Heisenberg magnet are completely decoupled. Because the classical
ground state does not have any intrinsic dynamics, it does not make
any sense to associate a spin current with it which would correspond
to moving magnetic moments.  Furthermore, if the classical Heisenberg
model is provided with Poisson bracket dynamics, the classical ground
state yields a stationary solution, since it minimizes the energy.
Clearly, such a completely stationary state cannot be used to
transport magnetization.  We conclude that for twisted spin
configurations Eq.~(\ref{eq:spincur1}) is not a physically meaningful
definition of the spin current operator.

\section{Effective spin currents with correct classical limit}

In order to arrive at a better definition, consider a non-equilibrium
situation, i.e.  start with a given density matrix at time $t=0$ and
let the system evolve according to the unitary dynamics generated by
the Hamiltonian in Eq.~(\ref{eq:Hamiltonian}). The equation of motion
(\ref{eq:eom}) then directly translates to a relation for the local
and instantaneous order parameter
\begin{equation}
  \partial_t  \langle {\bd{S}}_i \rangle_t 
  + {\bd{h}}_i \times \langle {\bd{S}}_i \rangle_t
  + \sum_j \langle {\bd I}_{i\to j}\rangle_t=0 \,.
  \label{eq:eomt}
\end{equation}
Here, $\langle \dots \rangle_t$ denotes an average with respect to the
time dependent density matrix.  It is then reasonable to demand that a
transport current can lead to an accumulation of magnetization, i.e. a
change in the magnitude of the local order parameter in time. For this
magnitude, we obtain the equation of motion
\begin{equation}
  \partial_t  |\langle {\bd{S}}_i \rangle_t| 
  + \sum_j {\hat{\bd m}}_i(t) 
  \cdot \langle{\bd I}_{i\to j}\rangle_t=0\,,
\end{equation}
where $\hat{\bd m}_i(t)=\langle{\bd S}_i\rangle_t/|\langle{\bd
  S}_i\rangle_t|$ is the time dependent direction of the order
parameter.  Note that only the longitudinal component of the naive
``current operator'' appears in this continuity equation without
source terms. It is precisely this contribution which we have
identified as the dominant one in our spin-wave calculation in
\cite{Schuetz03}.  The transverse components lead to a change in the
direction of the local order parameter, but they are largely
counteracted by the magnetic field term that acts as a source and
generates a precession.  If one wants to discuss the electric fields
generated by the magnetization dynamics, one either has to take into
account both the current ${\bd I}_{i\to j}$ and the local precessional
motion, or devise a way to make the cancelation explicit by including
part of the ``transverse current'' in an effective magnetic field. We
will attempt the second route here.

That this is a sensible way to proceed can be appreciated by a simple
approximate calculation. In the classical ground state, the
magnetization aligns parallel to the sum of the external and the
exchange field.  A necessary condition for the minimum of the
classical energy is the invariance under small variations of the
directions of the magnetization.  This leads to the condition
\cite{Schuetz03}
\begin{equation}
  {\bd{h}}_i^{\rm eff} \times 
  \langle {\bd{S}}_i \rangle 
  =0
  \; \; , \; \; 
  {\bd{h}}_i^{\rm eff} = 
  {\bd{h}}_i -  \sum_j J_{ij}  \langle {\bd{S}}_j \rangle
  \, .
  \label{eq:classical}
\end{equation} 
Note that the effective magnetic field contains a part of the exchange
interaction, which therefore should not be included into the
definition of the spin current operator to avoid double counting.
Rewriting the exact equation of motion (\ref{eq:eom}) in terms of the
effective magnetic field ${\bd{h}}_i^{\rm eff}$ defined in Eq.
(\ref{eq:classical}), we obtain
\begin{equation}
  \hbar \frac{ \partial   {\bd{S}}_i}{\partial t }    
  +   {\bd{h}}_i^{\rm eff}
  \times  {\bd{S}}_i +
  \sum_{j } {\bd{I}}^{\rm eff}_{ i \rightarrow j}   = 0
  \,,
  \label{eq:eom2}
\end{equation}
where
\begin{equation}
  {\bd{I}}^{\rm eff} _{ i \rightarrow j}  = 
  {J_{ij}} {\bd{S}}_i \times [  {\bd{S}}_j 
  - \langle {\bd{S}}_j \rangle  ]
  \label{eq:Ispin3}
  \; .
\end{equation}
Obviously,
\begin{equation}
  \langle {{\bd{I}}}^{\rm eff}_{i \rightarrow j} \rangle  =
  J_{ ij} \left[   \langle  {\bd{S}}_i \times  {\bd{S}}_j \rangle
    -   \langle  {\bd{S}}_i \rangle \times   
    \langle  {\bd{S}}_j \rangle
  \right] \; ,
  \label{eq:spincureff}
\end{equation}
which vanishes identically in the classical ground state, or if the
spins are treated within the mean-field approximation, where the spin
correlator is factorized.  Physically, this is due to the fact that
within the mean-field approximation the Heisenberg exchange
interaction is replaced by an effective magnetic field, so that the
different sites are uncorrelated and there are no degrees of freedom
to transfer magnetization between them. In this work we discuss only
localized spin models, so that charge degrees are not available to
transfer magnetization between different sites.  

\section{New definition of the spin current operator}

The definition of ${\bd I}^{\rm eff}_{i\to j}$ in
Eq.~(\ref{eq:Ispin3}) has the disadvantage of not being antisymmetric
with respect to the exchange of the site labels, although its
expectation value is obviously antisymmetric.  In order to cure this
problem and to generalize the concept of an effective current operator
beyond the mean-field description, we propose the following definition,
\begin{equation}
  \tilde{\bd I}_{i\to j} = {\bd I}_{i\to j}
  -{\bd \gamma}_{ij}({\bd \gamma}_{ij}\cdot{\bd I}_{i\to j})\,,
  \label{eq:Idef}
\end{equation}
with the unit vector
\begin{equation}
  {\bd \gamma}_{ij} = \frac{{\bd m}_i\times{\bd m}_j}
  {|{\bd m}_i\times{\bd m}_j|}\,.
\end{equation}
Thus, we interpret only the projection of ${\bd I}_{i\to j}$ onto the
plane spanned by the two local order parameters ${\bd m}_i$ and ${\bd
  m}_j$ as a physical transport current. The contribution subtracted
in Eq.~(\ref{eq:Idef}) can be incorporated in an effective magnetic
field. More precisely, the equilibrium expectation value of the exact
equation of motion (\ref{eq:eom}) can be rewritten as
\begin{equation}
  {\bd h}_i^{\rm eff}\times\langle{\bd S}_i\rangle 
  + \sum_j \langle\tilde{\bd I}_{i\to j}\rangle =0\,,
\end{equation}
with the effective magnetic field now defined as 
\begin{equation}
  {\bd h}^{\rm eff}_i = {\bd h}_i - \sum_j 
  \frac{\langle{\bd S}_i\times J_{ij}{\bd S}_j\rangle\cdot{\bd \gamma}_{ij}}
  {[\langle{\bd S}_i\rangle
    \times\langle{\bd S}_j\rangle]\cdot{\bd \gamma}_{ij}}
  \langle{\bd S}_j\rangle\,.
  \label{eq:heff}
\end{equation}
This reduces to Eq.~(\ref{eq:classical}) for the classical ground
state or at the mean-field level, where the correlation function in
the numerator is factorized.  The spin current operator defined in
Eq.~(\ref{eq:Idef}) is manifestly antisymmetric under the exchange of
the labels, as it should be. It implicitly depends on the spin
configuration via the unit vector ${\bd \gamma}_{ij}$, so that in
twisted spin configurations the spin current operator is a rather
complicated functional of the exchange couplings.  The fact that the
current operator of an interacting many body system is a complicated
functional of the interaction is well known from the theory of
interacting Fermi systems \cite{Pines89}. In particular, when the
effective interaction does not involve densities only the construction
of the current operator is not straightforward \cite{Metzner98}.

For explicit calculations we use a representation of $\tilde{\bd
  I}_{i\to j}$ in terms of spin operators quantized in local reference
frames with the z-axes pointing along $\hat{\bd m}_i$, i.e. we
decompose
\begin{equation}
  {\bd S}_i = \sum_{\alpha=1,2,3} S_i^{\alpha} {\bd e}_i^{\alpha}\,,
  \qquad
  S_i^{\alpha}={\bd e}_i^{\alpha}\cdot{\bd S}_i\,,
\end{equation}
with ${\bd e}_i^3= \hat{\bd{m}}_i$. One still has a freedom in the
orientation of the transverse basis $\{{\bd e}_i^1,{\bd e}_i^2\}$,
which can elegantly be parametrized, if one uses spherical basis
vectors ${\bd e}_i^{\pm}={\bd e}_i^1\pm i{\bd e}_i^2$. We can then
write
\begin{equation}
  {\bd e}_i^+ = e^{i\omega_{i\to j}} \tilde{\bd e}_i^+\,,
\end{equation}
where $\{\tilde{\bd e}_i^1,\tilde{\bd e}_i^2\}$ is the special
transverse basis where $\tilde{\bd e}_i^2=\tilde{\bd e}_j^2={\bd
  \gamma}_{ij}$. With this notation we obtain the following expression
for the spin current operator
\begin{eqnarray}
  \tilde{\bd I}_{i\to j} &=& \frac{J_{ij}}{2i}
  \Big[
  S_i^-S_j^+e^{i(\omega_{i\to j}-\omega_{j\to i})}
  \frac{\hat{\bd m}_i+\hat{\bd m}_j}2\nonumber\\
  &&-S_i^-S_j^-e^{i(\omega_{i\to j}+\omega_{j\to i})}
  \frac{\hat{\bd m}_i-\hat{\bd m}_j}2\nonumber\\
  &&+S_i^{\parallel}S_j^-e^{i\omega_{i\to j}}
  ({\bd \gamma}_{ij}\times\hat{\bd m}_i)\nonumber\\
  &&-S_i^-S_j^{\parallel}e^{i\omega_{j\to i}}
  ({\bd \gamma}_{ij}\times\hat{\bd m}_j)
  - {\rm H.c.}
  \Big]\,,
  \label{eq:Idecom}
\end{eqnarray}
where $S_i^{\pm}=S_i^1\pm i S_i^2={\bd e}^{\pm}_i\cdot{\bd S}_i$ are
the usual ladder operators and $S_i^{\parallel}=S_i^3=\hat{\bd
  m}_i\cdot{\bd S}_i$.  The third and fourth terms in this expression
couple longitudinal and transverse degrees of freedom and therefore do
not contribute to leading order in a spin-wave calculation. The first
and second summand are dominant for ferromagnetic and
antiferromagnetic rings respectively and have been discussed in detail
in \cite{Schuetz03} and \cite{Schuetz04}. For a magnetic field that
varies smoothly as one moves through the system, the magnetic moments
on neighboring lattice sites are almost collinear, so that in both
cases the component of the naive ``current operator'' ${\bd I}_{i\to
  j}$ along the local order parameter is the one that really
corresponds to the transport of magnetization.  In \cite{Schuetz03} we
had come to the same conclusion by invoking the gauge freedom in the
choice of the transverse axes of quantization. Rotating the coordinate
frame around $\hat{\bd m}_i$ corresponds to the gauge transformation
\begin{equation}
  \omega_{i\to j} \rightarrow \omega_{i\to j} + \alpha_i\,,
  \qquad 
  S_i^{\pm} \rightarrow S_i^{\pm} e^{\pm i\alpha_i}\,.
\end{equation}
By this gauge freedom one is then let to identify the derivative of
the Hamiltonian with respect to the gauge field $\omega_{i\to j}$ as
the relevant current operator.  A comparison with
Eq.~(\ref{eq:Idecom}) shows that this is indeed the longitudinal
component of ${\bd I}_{i\to j}$.  A more general gauge invariant
formulation of the Heisenberg model is discussed in \cite{Chandra90}
(see also \cite{Kopietz91}).  In these works, an $O(3)$ gauge field
${\bd{A}}_{i \rightarrow j}$ was introduced in a rather formal manner
to write the Heisenberg model in a gauge invariant way and to obtain
the spin stiffness tensor by means of differentiation with respect to
the gauge field \cite{Singh89}.

Note that the procedure adopted in this section is not restricted to
the isotropic Heisenberg interaction of the Hamiltonian
(\ref{eq:Hamiltonian}). For a general bilinear spin-spin interaction
of the form
\begin{equation}
  \hat{H} = \frac{1}{2} \sum_{i,j} {\bd{S}}_i \cdot {\bd J}_{ij}
  {\bd{S}}_j   - \sum_{i  }  {\bd{h}}_i \cdot
  {\bd{S}}_i
  \,,
\end{equation}
where ${\bd J}_{ij}$ is now a 3x3 matrix with ${\bf J}_{ij}^T={\bd
  J}_{ji}$, the equation of motion (\ref{eq:eom}) remains valid, if
the expression for ${\bd I}_{i\to j}$ is replaced by
\begin{equation}
  {\bd I}_{i\to j} = {\bd S}_i\times{\bd J}_{ij}{\bd S}_j\,.
\end{equation}
With this notation, Eqs.~(\ref{eq:Idef}-\ref{eq:heff}) still hold
(provided $J_{ij}$ is replaced by the matrix ${\bd J}_{ij}$) and part
of the naive current operator ${\bd I}_{i\to j}$ can again be absorbed
into the definition of an effective magnetic field.  

\section{Conclusion}

Let us emphasize again that our main point is rather simple: The
microscopic equation of motion (\ref{eq:eom}) contains only the
(lattice) divergence of the spin current operator, which is not
sufficient to fix its rotational part.  A certain part of the operator
$\sum_j J_{ij} {\bd{S}}_i \times {\bd{S}}_j$ leads to a
renormalization of the effective magnetic field and therefore should
not be included into the definition of the spin current operator, see
Eqs. (\ref{eq:classical}--\ref{eq:heff}).  The physical spin current,
which corresponds to the motion of magnetic dipoles, must be defined
such that a purely static twist in the ground state spin configuration
of a classical Heisenberg magnet is considered to be a renormalization
of the effective magnetic field, and does not contribute to the spin
current. To further substantiate our proposal for an effective current
operator, it would certainly be insightful to look for a more
microscopic derivation by starting from a model involving charge
degrees of freedom. It would also be instructive to explicitly
investigate non-equilibrium situations with time dependent
magnetizations.

We thank P. Bruno and K. Saito for discussions and for urging us to
clarify the proper definition of the spin current operator in magnetic
insulators.  This work was supported by the DFG via Forschergruppe FOR
412, Project No. KO 1442/5-3.


\begin{thebibliography}{99}
  
\bibitem{Schuetz03} F. Sch\"{u}tz, M. Kollar, and P. Kopietz, Phys.
  Rev. Lett. {\bf{91}}, 017205 (2003).
  
\bibitem{Schuetz04} F. Sch\"{u}tz, M. Kollar, and P. Kopietz, Phys.
  Rev.  B {\bf{69}}, 035313 (2004).
  
\bibitem{Rashba03} E. I. Rashba, Phys. Rev. B {\bf{68}}, 241315
  (2003); cond-mat/0404723; cond-mat/0408119.
  
\bibitem{Awschalom02}%
  See, for example, D.~D.\ Awschalom, D.\ Loss, and N.\ Samarth
  (Eds.), {\it{Semiconductor Spintronics and Quantum Computation}},
  (Springer, Berlin, 2002).

\bibitem{Meier02}%
  F.\ Meier and D.\ Loss, Phys.\ Rev.\ Lett.\ {\bf{90}}, 167204
  (2003).

\bibitem{Volovik87} G. E. Volovik, J. Phys. C {\bf 20}, L83 (1987).  
  
\bibitem{Pines89} D. Pines and P. Nozi\`{e}res, {\it{The Theory of
      Quantum Liquids}}, Volume I, (Addison-Wesley Advanced Book
  Classics, Redwood City, 1989).
  
\bibitem{Metzner98} W. Metzner, C. Castellani, and C. Di Castro, Adv.
  Phys. {\bf{47}}, 317 (1998).

\bibitem{Chandra90} P. Chandra, P. Coleman, and A. I. Larkin, J.
  Phys.: Condens. Matter {\bf{2}}, 7933 (1990).
  
\bibitem{Kopietz91} P. Kopietz and G. E. Castilla, Phys. Rev. B
  {\bf{43}}, 11100 (1991).
  
\bibitem{Singh89} R. R. P. Singh and D. Huse, Phys. Rev. B {\bf{40}},
  7247 (1989).
  
\end{thebibliography}
\end{document}